\documentclass[floatfix,twocolumn,citeautoscript,showpacs,amsmath,amssymb,epsf,aps,prb]{revtex4}
\usepackage{dcolumn}
\usepackage{epsfig}

\usepackage{graphicx}
\usepackage{longtable}
\usepackage{dcolumn}
\usepackage{bm}

\def\etal{{\em et al.}}
\def\trip_b{$^3$B$_1$ }
\def\sing_a{$^1$A$_1$ }
\def\af{$\alpha$ }

\def\Xa{X$\alpha$ }

\def\vr{\vec{r}}
\def\bo{\overline{\rho}^{\frac{1}{3}}}
\def\ro{\overline{\rho}}
\def\bt{\overline{\rho}^{\frac{2}{3}}}
\def\ss{\sigma}

\begin{document}

\title{ Accurate molecular energies by extrapolation of atomic energies using an analytic quantum 
mechanical model}

\author{Rajendra R. Zope}
\email{rzope@alchemy.nrl.navy.mil}
\affiliation{Department of Chemistry, George Washington University, Washington DC, 20052}

\author{Brett I. Dunlap}
\email{dunlap@nrl.navy.mil}
\affiliation{Code 6189, Theoretical Chemistry Section, US Naval Research Laboratory
Washington, DC 20375}

\date{\today}

\begin{abstract}

Using a new {\em analytic} quantum mechanical method based on 
Slater's X$\alpha$ method, we show that a fairly accurate estimate
of the total energy of a molecule can be obtained from the exact energies 
of its constituent atoms.
The mean absolute error in the total energies thus determined
for the G2 set of 56 molecules is about 16 kcal/mol, comparable to or better 
than some popular pure and hybrid density functional models.
\end{abstract}

\pacs{ }


\maketitle

    The total electronic energy is the central quantity in quantum mechanical 
methods that compute the ground state properties of many-electron systems.
These methods rely on the variational principle and can be broadly classified in 
two categories. The first category
is formed by {\it ab initio} quantum chemical methods which permit systematic 
improvement in accuracy and analytic computation of quantum mechanical 
matrix elements to machine precision\cite{Pople99}. The high computational cost of 
these methods restricts their application to those systems with a 
few atoms.  Density-functional-based computational models form the second
category. These models are based on the Kohn-Sham\cite{KS65} formulation of density 
functional theory (DFT) which offers an alternative way to determine 
ground state properties.  Today DFT based models have become the 
most popular choice for calculating these properties as they
provide results of sufficient accuracy at reduced computational costs.  
Unlike {\it ab initio} quantum chemical methods, however, the computational 
implementation of almost all DFT models require the use of numerical grids \cite{handy93}
making calculation of matrix elements at machine precision practically
impossible.  Round-off error, which grows as the square root of the number
of points, is eliminated by analytic methods \cite{Boys}.

   The desirable attribute of analytic computation, which is  accurate 
to machine precision, is also possible within 
DFT\cite{Cook95}.  This approach is based on variational and robust fits 
to the orbitals and the effective one-body Kohn-Sham potential using Gaussian 
basis sets\cite{Dunlap79}.
This procedure does not alter the density-functional one-particle equations, 
but makes its analytic solution possible.  The model has its roots in the early
density-functional model of Slater\cite{Slater51}, wherein the exchange 
potential $v_x$ is
proportional to the one-third power of the electron density $\rho$:
$$ v_x [\rho] =  - \alpha 
\frac{3}{2}
 \Bigl( \frac{3}{\pi} \Bigr )^{1/3} \rho^{1/3}(\vec{r}),$$
where $\alpha$ is called the Slater exchange parameter.
We extended this analytic density-functional model so that an
atom-dependent exchange parameter $\alpha,$ which scales the exchange 
potential differently for each element, can be used in heteroatomic molecules 
or solids\cite{Dunlap03}.
Early numerical attempts to use atom dependent \af values had 
a mathematically undefined total energy that results 
from the discontinuity in the potential 
at the boundary of the muffin-tin sphere that enclosed atoms or ions,
in the Multiple-Scattering \Xa method\cite{MT}.
In our method the total energy is both well defined and stationary with 
respect to variations of {\it all} linear combination of atomic-orbital (LCAO) and 
Kohn-Sham-potential
fitting  coefficients.
The energy is a function of basis sets and Slater's $\alpha$ for each element,
\begin{eqnarray}
  E(\{\alpha_i\}) &  = & \sum_i <\phi_i| f_1| \phi_i> + 2 \langle \rho \vert\vert\ro\rangle
                - \langle \ro\vert\vert\ro\rangle    \nonumber \\
          & & \,\, - \sum_{\sigma = \uparrow, \downarrow} \Biggl [
            \frac{4}{3} \langle g_{\ss}  \, \bo_{\!\ss} \rangle
           - \frac{2}{3} \langle {\bo}_{\!\ss} \,  {\bo}_{\!\ss}  \, {\bt}_{\!\ss} \rangle  \nonumber \\
         & & \,\, \, +  \,\frac{1}{3} \langle {\bt}_{\!\ss} \,  \, {\bt}_{\!\ss} \rangle \Biggr ].
    \label{eq:1}
\end{eqnarray}
  Here, $f_1$ is the one-electron part of the Hamiltonian,
$\rho_{\sigma}$ is the spin density and an overbar represents an LCAO fit, 
and $ g_{\ss}$, the partitioned $3/4$ power of the exchange energy density:
\begin{equation}
  g_{\ss} (\vr) = \sum_{ij} \alpha(i) \, \alpha(j) \, D_{ij}^{\ss} (\vr),
\end{equation}
where $D_{ij}^{\ss} (\vr)$ is the diagonal part of the spin density matrix, and the function,
\begin{equation}
   \alpha(i) = \Bigl [ 3 \, \alpha_i \, \bigl ( \frac{3}{4\pi} \bigr )^{\frac{1}{3}} \bigr ) \Bigr  ]^{3/8}
\end{equation}
contains $\alpha_i$, the \af within the muffin-tin in the \Xa method, for
the atom on which the atomic orbital $i$ is centered.  This unique expression for
the total energy is obtained by adding different fitted expressions for the same components
of the energy so as to cancel all first-order errors in the energy due to all fits.
All orbitals and fits are determined through variation of Eq. ~\ref{eq:1}.

    This method inherits from Slater's  \Xa method the physically appealing
advantage, which it had over all other quantum-chemical methods, that atoms
dissociate correctly.  For these molecules and this range
of $\alpha$ values all dissociated atoms turn out to be neutral in spin-polarized (high-spin),
broken-symmetry (the highest symmetry that gives integral occupation numbers in
fractional-occupation-number) calculations.  Using Eq. 1 and atomic $\alpha$ values
is a single approximation,
like the local density approximation is a single approximations and not a different parameterization
for each range of densities.
In particular, we determine a set of atomic \af values that yield the exact atomic (EA) energies
and use them  to compute total energies of the G2 set\cite{Becke93} of 56 molecules. 

%
\begin{table}
\begin{ruledtabular}
\caption{The optimal $\alpha$ values that yield the
{\em exact} atomic energies in the highest symmetry for which the solutions
have integral occupation numbers for the analytic DFT calculations. 
The {\em exact} atomic energies given in the last column are from Ref.\, \onlinecite{clemmenti}.
The basis sets are I: 6311G**/RI-J, II: DZVP/A2.}
\label{table:aalpha}
\begin{tabular}{llllr}
   &  Basis I     & Basis II   &   Numerical &  E (a.u.)      \\
\hline
  H &   0.77739     &   0.78124 &   0.77679  &     -0.500     \\
 Li &   0.79169     &   0.79211 &   0.79118  &     -7.478     \\
 Be &   0.79574     &   0.79614 &   0.79526  &    -14.667     \\
  B &   0.78675     &   0.78677 &   0.78744  &    -24.654     \\
  C &   0.77677     &   0.77665 &   0.77657  &    -37.845     \\
  N &   0.76747     &   0.76726 &   0.76654  &    -54.590     \\
  O &   0.76500     &   0.76448 &   0.76454  &    -75.067     \\
  F &   0.76066     &   0.76001 &   0.75954  &    -99.731     \\
 Na &   0.75204     &   0.75287 &   0.75110  &   -162.260     \\
 Mg &   0.74994     &   0.75120 &   0.74942  &   -200.060     \\
 Al &   0.74822     &   0.74869 &   0.74797  &   -242.370     \\
 Si &   0.74539     &   0.74602 &   0.74521  &   -289.370     \\
  P &   0.74324     &   0.74397 &   0.74309  &   -341.270     \\
  S &   0.74262     &   0.74350 &   0.74270  &   -398.140     \\
 Cl &   0.74197     &   0.74272 &   0.74183  &   -460.200     \\
\end{tabular}
\end{ruledtabular}
\end{table}

    Our calculations use Gaussian basis sets to fit both the orbitals
and the Kohn-Sham potential. Here, we choose two different combinations of 
basis sets. For orbitals, we use the valence triple-${\zeta}$ (TZ) 6-311G**
basis\cite{O1} and the DGauss valence double-$\zeta$ basis\cite{O2}
 set (DZVP2). The s-type fitting bases are obtained by scaling the exponents 
of the $s$ part of the orbital basis,
by two to fit $\ro$, by $2/3$ to fit $\bo$, and by $4/3$ to fit $\bt$.
For the non-zero angular-momentum components the resolution-of-the-identity-J 
(RI-J)\cite{EWTR97} and A2 \cite{O2} basis sets 
are used without scaling to fit the Coulomb potential as well as the exchange-correlation
 part of the Kohn-Sham potential.  Thus, two combination 
sets  6-311G**/RI-J and  DZVP/A2 of bases were used for obtaining the atomic 
\af values.  The \af optimization was performed using the PERL scripts.

We determine of \af values that give the
{\em exact} total electronic energy for each atom in the molecule.
That set of \af values is given  in Table \ref{table:aalpha} for 
each basis set.  
The ``exact" atomic energies\cite{clemmenti} that these
values of ${\alpha}$ reproduce are also included in the same 
table. These \af values are obtained by the Newton-Raphson procedure 
to zero the function
 $ f(\alpha) =  E (\alpha) - E_{exact},  $ 
where $E(\alpha)$ and $ E_{exact}$ are the self-consistent atomic energy for each
value of \af and the {\em exact} total 
energies, respectively. 
The optimal $\alpha$ values are obtained for {\em exact} atomic energies in the 
highest symmetry for which the solutions have integral occupation numbers. 
The third set of \af values is  obtained from a spherically-symmetric, numerical 
electronic-structure code for atoms. 
The \af values for the DZVP2/A2 basis are usually larger than 
those for the 6-311G**/RI-J as the orbital variational principle requires lower 
energy from a larger basis set and increasing \af lowers the total energy.
The fitting basis has a smaller effect on the total energies than the orbital
basis\cite{Dunlap03}.

\begin{table}     
\begin{ruledtabular}    
\caption{Absolute Error in total molecular energies with respect to its {\em exact} 
value for the G2 set of molecules.
The computational models are-
 M1:  present/6311G**/RI-J, 
 M2:  PBE/6311G**
 M3:  B3LYP/6311G*. 
The absolute errors are  at the optimized geomertries in 
the repsective model. Last column contains ``exact" total energy. 
The errors and mean absolute error (MAE) are in kcal/mol. (See text for 
more details).}
\label{table:error}
\begin{tabular}{lcccc}
\hline  
         &         M1        &   M2      &   M3         &   Exact (a.u.) \\
 H$_2$ 	 &  	  1.83 &  0.44 &  0.20 &   -1.17530  \\
 LiH 	 &  	  1.44 &  1.17 &  0.75 &   -8.06995  \\
 BeH 	 &  	  0.64 &  1.43 &  0.88 &  -15.24604  \\
 CH 	 &  	  1.23 &  2.48 &  0.63 &  -38.47838  \\
 CH$_2$($^3B_1$) 	 &  	  0.40 &  2.20 &  0.78 &  -39.14747  \\
 CH$_2$($^1A_1$) 	 &  	  1.64 &  2.75 &  0.54 &  -39.13265  \\
 CH$_3$ 	 &  	  0.53 &  2.32 &  0.94 &  -39.83328  \\
 CH$_4$ 	 &  	  0.87 &  2.46 &  0.96 &  -40.51288  \\
 NH 	 &  	  1.20 &  2.76 &  0.65 &  -55.22291  \\
 NH$_2$ 	 &  	  1.79 &  2.78 &  0.73 &  -55.87924  \\
 NH$_3$ 	 &  	  1.87 &  3.03 &  0.57 &  -56.56378  \\
 OH 	 &  	  0.56 &  3.19 &  0.84 &  -75.73640  \\
 H$_2$O 	 &  	  0.30 &  3.53 &  0.49 &  -76.43687  \\
 HF 	 &  	  0.30 &  3.96 &  0.67 &  -100.45522  \\
 Li$_2$ 	 &  	  1.31 &  1.96 &  0.96 &  -14.99488  \\
 LiF 	 &  	  0.65 &  4.65 &  1.30 &  -107.43019  \\
 C$_2$H$_2$ 	 &  	  1.29 &  4.34 &  0.87 &  -77.33589  \\
 C$_2$H$_4$ 	 &  	  0.67 &  4.38 &  1.28 &  -78.58624  \\
 C$_2$H$_6$ 	 &  	  0.08 &  4.56 &  1.55 &  -79.82257  \\
 CN 	 &  	  0.84 &  4.13 &  0.77 &  -92.72025  \\
 HCN 	 &  	  0.10 &  4.79 &  0.60 &  -93.43906  \\
 CO 	 &  	  1.77 &  4.93 &  0.98 &  -113.32506  \\
 HCO 	 &  	  2.14 &  4.49 &  1.43 &  -113.85550  \\
 H$_2$CO  	 &  	  1.55 &  4.85 &  1.35 &  -114.50705  \\
 H$_3$COH  	 &  	  0.75 &  5.36 &  1.39 &  -115.72729  \\
 N$_2$ 	 &  	  0.99 &  5.02 &  0.54 &  -109.54414  \\
 N$_2$H$_4$ 	 &  	  1.78 &  5.34 &  1.10 &  -111.87768  \\
 NO 	 &  	  0.76 &  4.85 &  1.20 &  -129.90066  \\
 O$_2$ 	 &  	  2.73 &  4.73 &  1.80 &  -150.32587  \\
 H$_2$O$_2$ 	 &  	  1.08 &  5.79 &  1.38 &  -151.56204  \\
 F$_2$ 	 &  	  2.06 &  6.38 &  1.83 &  -199.52335  \\
 CO$_2$ 	 &  	  4.97 &  6.81 &  1.96 &  -188.59875  \\
 SiH$_2$($^1A_1$) 	 &  	  1.91 &  7.13 &  1.32 &  -290.61127  \\
 SiH$_2$($^3B_1$) 	 &  	  0.70 &  6.76 &  1.37 &  -290.57828  \\
 SiH$_3$ 	 &  	  2.35 &  7.20 &  1.38 &  -291.23127  \\
 SiH$_4$ 	 &  	  2.92 &  7.51 &  1.47 &  -291.88218  \\
 PH$_2$ 	 &  	  1.96 &  7.76 &  0.95 &  -342.51350  \\
 PH$_3$ 	 &  	  2.84 &  8.13 &  0.79 &  -343.15565  \\
 H$_2$S 	 &  	  1.37 &  9.56 &  0.39 &  -399.43051  \\
 HCl 	 &  	  0.36 & 11.31 &  1.65 &  -460.86924  \\
 Na$_2$ 	 &  	  0.87 &  9.52 &  2.47 &  -324.54677  \\
 Si$_2$ 	 &  	  0.19 & 13.29 &  1.82 &  -578.85904  \\
 P$_2$ 	 &  	  1.70 & 15.59 &  0.33 &  -682.72677  \\
 S$_2$ 	 &  	  0.63 & 18.37 &  1.35 &  -796.44191  \\
 Cl$_2$ 	 &  	  0.48 & 22.60 &  4.00 &  -920.49227  \\
 NaCl 	 &  	  0.66 & 16.28 &  0.72 &  -622.61585  \\
 SiO 	 &  	  0.57 & 10.02 &  1.10 &  -364.74170  \\
 CS 	 &  	  0.59 & 11.42 &  0.54 &  -436.25782  \\
 SO 	 &  	  1.28 & 12.18 &  0.32 &  -473.40636  \\
 ClO 	 &  	  1.12 & 14.00 &  1.39 &  -535.36947  \\
 ClF 	 &  	  1.41 & 14.74 &  1.25 &  -560.02885  \\
 Si$_2$H$_6$ 	 &  	  3.98 & 14.51 &  2.52 &  -582.58381  \\
 CH$_3$Cl  	 &  	  0.50 & 13.37 &  0.99 &  -500.17224  \\
 H$_3$CSH     	 &  	  0.36 & 11.65 &  0.20 &  -438.73830  \\
 HOCl 	 &  	  0.75 & 14.24 &  1.35 &  -536.02883  \\
 SO$_2$ 	 &  	  1.68 & 16.03 &  1.38 &  -548.68595  \\
\hline  
  MAE  & 17.3 &   100.8 &  15.3 &  - \\
 \end{tabular}         
\end{ruledtabular}    
\end{table}     

  The second step consists of computing total energy of molecules using Eq. 1 
and these \af values, which in effect extrapolates the atomic energies.
All molecules belonging to the G2 set are optimized and the absolute 
errors in the total energies of these molecules are computed. These are 
given in Table ~\ref{table:error}.  We choose the G2 set
as accurate experimental atomization energies and thus total energies 
for these molecules are well known. 
We used the zero-point corrected atomization energies from Ref.\, \onlinecite{VSXC} and
the atomic energies from Table I to synthesize an experimental total energy.
The G2 set is
routinely used for performance appraisal of density functional models.
While our energy is not obviously a density-functional, it shares a common
root \cite{Slater51} with DFT.  Of all methods with this common root,
only one other \cite{Cook95} has been treated analytically in these fifty years.
The errors in the total energies computed in our model are compared with the
popular density functional and hybrid models in Table ~\ref{table:MAE}. 
These density functional models are certainly more sophisticated intrinsically 
than either the G\'asp\'ar-Kohn-Sham (GKS) or \Xa methods. The total energies
corresponding to these sophisticated functionals, however, are not obtained in
particularly sophisticated fashion.  It is necessary to use a numerical
grid for these more sophisticated functionals.  Consequently 
the energies are dependent on how one (here GAUSSIAN03\cite{GAUSSIAN03})
chooses to orient each molecule relative to that necessary grid. On the other hand,
energies obtained in our model use a less sophisticated functional, but are
independent of molecular orientation, and thus could be 
obtained to whatever accuracy is needed to judge between analytic DFT models.
The sophisticated pure and hybrid density functional models that we
compare with are: the parameter-free PBE\cite{PBE} and the three-parameter
hybrid density functional B3LYP, which mixes 
Hartree-Fock (HF) exchange with the local exchange functional and 
Becke's generalized gradient exchange functional\cite{Becke93} along with 
local correlation\cite{VWN80} and the LYP\cite{LYP88} correlation functionals. 
The mixing coefficients were empirically determined to minimize atomization energies,
not total (less zero-point) energies addressed in this work.
We also computed the total 
energies for several other  density functional models. These include the local
density approximation (LDA)\cite{VWN80}, BLYP-GGA\cite{Becke93,LYP88},
Perdew-Wang (PW91) GGA\cite{PW91}, the meta-GGA functional
containing kinetic densities due to Voorhis and Scuseria (VSXC)\cite{VSXC},
the empirical GGA due to Handy and Cohen (HCTH407)\cite{HCTH407}, and 
the hybrid PBE1PBE (also called PBE0) functional\cite{BEP97}. 
The HCTH407 functional contains 15 parameters that are fitted to a database 
of 407 properties including total energies. The PBE1PBE model is a 
hybrid functional of PBE and 25\% of HF exchange.
The total energies used for comparison 
are computed with the 6-311G** basis set by the GAUSSIAN03 code at fully optimized 
geometries within each respective model. The default fine mesh is used in all calculation.

  It is well known that Gaussian are not ideal for computing atomic energies, and therefore
larger Gaussian basis sets must be used. Thus we also calculated total energies for the 
PBE models at the geometries optimized by the NRLMOL code\cite{NRLMOL}
using the NRLMOL basis\cite{NRLMOL-basis}.  The NRLMOL Gaussian basis set is
optimized for the PBE density functional model and is much larger than 
the 6-311G** basis. These results are also included in Table \ref{table:MAE}. 
The mean G2-set errors are also tabulated as a test for this, and potentially
other, systematic errors.
As the mean errors are small, Gaussian basis set incompleteness in computing
atomic energies is not an important consideration.

In this study of extrapolating atomic energies only the MAE in total 
energy for the density functional models in Table \ref{table:MAE} is of concern.
The zero-point energy is not included in total energies as the analytic second derivative 
calculation required for vibrational frequencies is not yet available in our model.
Correcting for zero-point motion can introduce a small error particularly 
for polyatomic systems. 
The MAE in total energy obtained by our method is 16.2 kcal/mol for the 
DZVP2/A2  and 17.3 kcal/mol for the 6-311G**/RI-J basis.  The comparison of our 
MAE with that obtained from the most widely used pure and hybrid density-functional 
models is favorable.
\begin{table}
\begin{ruledtabular}
\caption{Mean absolute error (MAE) and mean error (ME) in total energies of G2 
set of 56 molecules for different models. The errors 
are in kcal/mol and are at optimized geometries in respective model.
M1:  present/6-311G**/RI-J, M2:  present/DGDZVP2/A2 (see text for more details).}
 \label{table:MAE}
\begin{tabular}{lrdl}
  Model &   MAE  &  ME \\
 \hline
  present (M1)    &   17  & -0.4 \\
  present (M2)    &   16  & -0.5 \\
  LDA             &   532 &  20.0 \\
  PBE-GGA         &    101 &   3.9  \\
  BLYP-GGA        &    13 &   0.4  \\
  PW91-GGA        &    19 &   3.8   \\
  PBE-GGA (NRLMOL)&    87 &   3.1    \\
  HCTH407-GGA     &    14 &   0.3    \\
  VSXC-meta GGA   &    60 &  -1.7    \\
  B3LYP-hybrid GGA &    15 &  0.3   \\
  PBE1PBE-hybrid GGA&    91 &  3.9   \\
\end{tabular}
\end{ruledtabular}
\end{table}
This simple way of determining total energies of molecules by extrapolation is remarkably
accurate.  It is also likely to be significantly more accurate than the {\it ab initio}
methods that can treat the largest G2 molecules today because the HF energy
bounds the exact energy from above.

Our model gives total energies comparable to the B3LYP, which is widely accepted as the
most accurate hybrid density-functional model, because   
it gives the best atomization energies.
The B3LYP/6-311G** MAE in atomization energy (for the G2 set) is about 4.5 kcal/mol.
Using the larger 6-311+G(3df,2p) basis set the B3LYP MAE in atomization energy drops 
to 2.27 kcal/mol.
As the error in the total atomic energies in the present model is zero 
by construction, its error in atomization energy is the same as its error 
in the total energy.
For B3LYP, and perhaps other models, this essentially single experimental quantity,
may be computed in ways that can differ by almost an order of magnitude in accuracy.
Perhaps, density-functional 
models must benefit from cancellation of errors in total energies of atoms 
and molecules to give atomization energies comparable to or better than
{\it ab initio} methods. 
Unexpectedly, these models provide quite accurate total energies as well
as atomization energies.

The mixing parameters in the hybrid B3LYP method are empirically obtained 
by minimizing the errors in atomization energies. 
Like B3LYP, our method can also be parameterized for atomization 
energies by optimizing the MAE in atomization 
energies of the G2 set of molecules.  Such an optimized parametrization gives MAE 
in atomization energy that is intermediate between the
PBE-GGA and B3LYP. 
The calculations could be further improved in several ways. 
 A larger basis also lowers the B3LYP error\cite{VSXC}. A larger basis
might lower our MAE, but the basis-set effect
is largely canceled by adjusting \af to get exact atomic energies, and
high angular-momentum functions are less important for HF-less functionals.
Perhaps the \af's should be adjusted towards the GKS value for valence
or open-shell electrons.  That seems
algorithmically possible, but departs from Slater's original model for the
binding of atoms in molecules and solids.

      In conclusion, we have shown that fairly accurate estimates of the total 
energy of molecules can be obtained by extrapolation using experimental atomic
energies and an analytic quantum mechanical model. The computational model is illustrated
by computing the geometry-optimized total electronic energy of the G2 set of 56
molecules, without grids and thus with machine-precision
matrix elements.  It is shown to perform as well as or better than some widely used
pure and hybrid density-functional models in computing total molecular electronic
energies simply by extrapolating atomic energies quantum mechanically, according to
Slater's prescription.


        The Office of Naval Research, directly and through the Naval Research Laboratory, and and the 
Department of Defense's   High Performance Computing Modernization Program, through the 
CHSSI
Project MBD-5, supported this work.
The calculations were performed at the Army Research Laboratory 
Major Shared Resource Center (ARL MSRC).


\begin{thebibliography}{99}
 \bibitem{Pople99}
 J. A. Pople, Rev. Mod. Phys.  {\bf 71}, 1267 (1999).


\bibitem{KS65} W. Kohn and L. J. Sham, Phys. Rev.  {\bf 140},  A1133 (1965).


\bibitem{handy93} 
C. W. Murray, N. C. Handy, and G. J. Laming, Mol. Phys.  {\bf 78},  997 (1993).

\bibitem{Boys}
S. F. Boys, Proc. Roy. Soc. A {\bf 200}, 542 (1950).

 \bibitem{Cook95}
 K. S. Werpetinski and M. Cook,  Phys. Rev. A   {\bf 52}, 3397 (1995);
 J. Chem. Phys.   {\bf 106}, 7124 (1997).

\bibitem{Dunlap79} B. I. Dunlap, J. W. D. Connolly, and J. R. Sabin, J. Chem. Phys. 
{\bf 71}, 3396 (1979); B.I. Dunlap, Phys. Chem. Chem. Phys. {\bf 2}, 2113 (2000).

\bibitem{Slater51} 
 J. C. Slater,  Phys. Rev. {\bf 81}, 385 (1951).

\bibitem{Dunlap03}  B. I. Dunlap, J. Phys. Chem. {\bf 107} 10082 (2003).

\bibitem{MT} 
 Connolly, J. W. D. In {\it  Modern Theoretical Chemistry;} Segal, G. A., Ed.; Plenum: New York, 1977; Vol. 7, p 105. 

\bibitem{PBE}
  J. P. Perdew, K. Burke and M. Ernzerhof,  Phys.  Rev.  Lett.   {\bf  77 }, 3865(1996);
       {\bf 78}, 1396 (1997)(E).

\bibitem{Becke93} A. D. Becke, J. Chem. Phys.  {\bf 98} 1372  (1993).

\bibitem{LYP88} C. Lee, W. Yang, and R. G. Parr,  Phys. Rev. B   {\bf 37} 785  (1988).

\bibitem{O1}
 R. Krishnan, J. S. Binkley, R. Seeger, and J. A. Pople, J. Chem. Phys. {\bf 72}, 650, (1980);
 McLean, A. D.; Chandler, G. S. J. Chem. Phys. {\bf 72}, 5639 (1980).

\bibitem{O2}  J. Andzelm, E. Wimmer, J. Phys. B, 172, 307 (1991); J. Chem. Phys. {\bf 96}, 1280  (1992);
N. Godbout, D. R. Salahub, J. Andzelm, E. Wimmer, Can. J. Chem. {\bf 70}, 560  (1992).

\bibitem{EWTR97}  K. Eichkorn, F. Weigend, O. Treutler, and  R. Ahlrichs, Theor. Chem. Acc. {\bf 97} (1997) 119.


\bibitem{clemmenti}  A. Veillard and E. Clementi, J. Chem. Phys. {\bf 49}, 2415 (1968).

\bibitem{VWN80}  S. H. Vosko, L. Wilk, and M. Nusair, Can. J. Phys. {\bf 58}, 1200 (1980).

\bibitem{PW91}  
J. Perdew \etal, Phys. Rev. B {\bf 46}, 6671 (1992); {\bf 48}, 4978(E) (1993).

\bibitem{VSXC}  
T. Van Voorhis and G. E. Scuseria, J. Chem. Phys. {\bf 109}, 400 (1998).

\bibitem{HCTH407}  
A. D. Boese and N. C. Handy, J. Chem. Phys. {\bf 115}, 5497 (2001).

\bibitem{BEP97}  
K. Burke, M. Ernzerhof and J. P. Perdew  Chem. Phys. Lett. {\bf 265}, 115 (1997).

\bibitem{GAUSSIAN03}  
 M. J. Frisch \etal, Gaussian, Inc., Wallingford CT, 2004.

\bibitem{NRLMOL}  
 M. R. Pederson and K. A. Jackson, Phys. Rev. B. {\bf 41}, 7453 (1990);
{\bf 43}, 7312 (1991); K. A. Jackson and M. R. Pederson, {\it ibid}
{\bf 42}, 3276 (1990).

\bibitem{NRLMOL-basis}  
 D. V. Porezag and M. R. Pederson, Phys. Rev. B. {\bf 54}, 7830 (1996);
 D. V. Porezag, PhD thesis: http://archiv.tu-chemnitz.de/pub/1997/0025.


\end{thebibliography}
\end{document}